\documentclass[number,preprint]{elsarticle}
\usepackage{longtable}
\usepackage{graphicx}
\pdfoutput=1
\setlongtables
\journal{Acta Astronautica}

\begin{document}

\begin{frontmatter}

\title{\huge\bf The Edge of Space: Revisiting the Karman Line}

\author[cfa]{Jonathan C. McDowell}
\ead{jcm@cfa.harvard.edu}

\address[cfa]{Harvard-Smithsonian Center for Astrophysics, 60 Garden St, Cambridge, MA 02138, USA}

\date{2018 Jul 6}

\begin{abstract}

In this paper I revisit proposed definitions of the boundary between the
Earth's atmosphere and outer space, considering orbital and suborbital
trajectories used by space vehicles. In particular, I investigate the
inner edge of outer space from historical, physical and technological viewpoints and
propose 80 kilometers as a more appropriate boundary than the currently
popular 100 km Von K\'arm\'an line.

\end{abstract}

\begin{keyword}
astronautics \sep Karman line \sep atmosphere \sep mesosphere \sep perigee \sep boundaries
\end{keyword}

\end{frontmatter}
\clearpage

\section{The Edge of Space}

\subsection{Introduction}\label{Sec101}

The argument about where the atmosphere ends and space begins predates the
launch of the first Sputnik (e.g. \cite{Jastrow1957}). The most widely - but
not universally - accepted boundary is the so-called Karman Line, nowadays
usually set to be 100 km altitude, but boundaries ranging from 30 km to 1.5 million
km have been suggested, as summarized in a 1996 book by Goedhart \cite{Goedhart1996}. 

Although the subject has not been much addressed in the physics
literature, there is an extensive law/policy literature on the subject - 
see e.g. \cite{Haley1963} \cite{HarrisHarris2006}, 
\cite{Monahan2008}, \cite{Neto2015} \cite{Hansen2015}.
Hansen \cite{Hansen2015} notes that COPUOS has wrestled with the issue continuously since
1966 \cite{COPUOS2002} without a conclusion. COPUOS, the Committee on Peaceful Uses Of Outer Space,
was established in 1959 and is the UN body dealing with astronautics. In COPUOS the USSR
repeatedly proposed either 100 or 110 km but the US rejected any definition.

As early as 1957 Robert Jastrow (\cite{Jastrow1957}, cited in \cite{Neto2015}) suggested
that the air space boundary should be at 100 km.
Goedhart (p. 3) lists almost 30 different proposals from the 1951-1962 period
for an altitude boundary ranging from 20 to 400 km; most values are in the 75-100 km
range. A number of these authors suggest that the large variations with time of
atmospheric properties make it futile to locate a true boundary of space
based on physical arguments. In this paper I will argue the contrary: there
is a moderately-well-defined boundary of space, it coincides with the Karman line
as originally defined, and that line is close to 80 km, not 100 km.

\subsection{The Functionalist Objection}\label{Sec102}

There have been objections (particularly in the United States) to defining
any {\it legal}~boundary of space on the grounds that it could cause disputes about airspace
violations below the boundary, or that too high a boundary could inhibit
future space activities. Those advocating this position, beginning with
McDougal and Lipson \cite{McDougalLipson1958}, are sometimes
referred to as `functionalists' (see also \cite{Neto2015},\cite{Monahan2008}).
The functionalist approach would ensure that long range ballistic missiles 
were not made subject to international agreements on `space objects', which may explain
part of its appeal to the US establishment.

The general tenor of these objections, however, seem applicable to any
law about anything. Functionalists also suggest that space law would
apply to an orbital rocket even while it was within the atmosphere, or
possibly on the ground. This seems unnecessary as national and
international law would already apply.  Suggestions that the purpose of
a vehicle, not its location, should determine the legal regime may be
appropriate for questions of licensing, but will not help if a vehicle
classified as belonging to one regime collides or interferes with one
from another regime. 

The special need for distinct laws specifically for space (and thus the need
for legal definition of space) arises from:
\begin{itemize}
\item The lack of national boundaries in space (analogous to international waters)
\item Objects in space may remain in motion relative to the Earth for long periods of time
(depending on the orbit, from days to millenia) without the need to refuel or land.
\item The large area swept out by a space object in a given time
due to the large kinetic energies involved in space travel, meaning that
a given space activity will extend over a wide area rather than 
be spatially localized as most Earth-based activities are.
\item The high destructive potential of collisions, since their effects also are felt
over a wide area.
\end{itemize}
The latter two considerations in particular are specific to the region above
the atmosphere where orbital dynamics dominate, although they do not apply
to activities on the surfaces or in the atmospheres of other worlds such as the Moon or Mars.

In any case, my main interest here is to define a boundary of space for the use of historians of
spaceflight, rather than to define a legal regime, so I will give no further
consideration to the functionalist view. To answer questions such as `how many astronauts
have flown in space?' or `how many European Space Agency rockets have reached space?', we need
to adopt a definition of space, even if it is not a legal one.

I do not argue that there needs to be a single definition of `space' that applies
in all contexts. Physicists, lawyers and historians may need a boundary of space
for different purposes and to address different questions; the 'edge of space' might 
be defined differently in different international fora. Nevertheless it is useful
for those definitions to be based on a common and accurate understanding of the physical 
conditions at the air-space boundary, and I hope that this article can make a positive
contribution in that respect.

\subsection{McDowell (1994) proposal}\label{Sec103}

Milt Thompson's book \cite{Thompson1992}
talks about his flights in the X-15 rocketplane to the `edge of space'. In my Quest
article `The X-15 Spaceplane' \cite{McDowell1994} I discussed this term
and concluded that the correct choice for the edge was 80 km, not 100 km. 
The discussion is repeated and expanded in this paper. In part, I argued:

\begin{quote}
{\it
In the late 1950s the USAF decided to award `astronaut wings' to pilots
flying above 50 statute miles. This boundary was chosen as a nice round 
figure, but I want to argue that it is also the right choice from a
physical point of view ....  
... it seems natural to choose 
the outermost [physical atmospheric] boundary,
the mesopause, as the physical boundary which marks the edge of space.
It turns out that the traditional value for the height of the mesopause, 
80 km, is also within 500 metres of the 50 mile
`astronaut wings' boundary historically used by the USAF.
I therefore suggest that we adopt as the formal boundary of space an
altitude of exactly 80 km, representing the typical location of the
mesopause.}
\end{quote}

In this paper I expand on the arguments in my 1994 article. 
All references to altitudes are intended to represent geodetic height
(height above reference ellipsoid); I ignore local topography.
Note that in some astronautical contexts use is made of a `geocentric' height
relative to a fictional spherical Earth, which may differ by of order 10 km.

\section{Cultural arguments: historical definitions of the edge of space}\label{Sec2}

In recent decades the 100 km Karman line has gained ascendancy as the most
commonly used boundary, notably for the Ansari X-Prize won by the Spaceship One
team.

The `official' status of the von Karman line, such as it is, comes from the undated
paper `100 km Altitude Boundary For Astronautics' \cite{FAI2004}
on the web site of the Astronautics Records Commission (ICARE) of the Fédération Aéronautique Internationale 
(FAI), which certifies world records for aeronautics and astronautics. It is unfortunate that discussion
in this official document in the section `demonstration of usefulness of Karman line'
appears to be poorly researched:

\begin{quote}

{\it In the early 1960s the U.S. X-15 Aircraft was flown up to 108 km. In
that part of the flight it was really a free falling rocket, with no
aerodynamic control possible. In fact, it was considered an
astronautical flight, and the pilot got, as a consequence, his
astronautical wings, i.e. the recognition of being an astronaut.}
\end{quote}

This is not incorrect, but the USAF considered all X-15 flights above 80 km
as astronautical flights and gave those pilots astronaut wings.
So this paragraph would argue for 80 km, not 100 km.
The first non-NASA pilot to be awarded astronaut wings was X-15 pilot
R. White, as described in Life Magazine (Aug 3 1962):
\begin{quote}
Major Bob White of the US Air Force is the nation's newest space hero. [...] He has [...]
a brand-new award on his chest that makes him a member of the nation's
most exclusive club. It was a special set of pilot's wings that signified he
had flown higher than 50 miles above the earth and thereby had qualified as
a spaceman.
\end{quote}

White flew to 95 km on 1962 Jul 17. If a limit of 100 km instead of 80
km is used, White, Robert Rushworth, Jack McKay, Bill Dana and Mike
Adams lose their space traveler status (Joe Engle keeps his because he
later flew on Shuttle, and Joe Walker passed the Karman line on his X-15
flights).

In the past few years, even lower values have been proposed. The prominent
astrophysicist Alan Stern
has argued (personal communication) for balloon altitudes in the 30-35 km range as
being `space' or `near space'; Stern is involved in
the World View high-altitude-balloon near-space tourism venture.

\section{Physical boundaries in the atmosphere}{\label{Sec3}}

\subsection{Atmospheric layers: the mesopause as a proposed boundary}\label{Sec301}

As you leave the surface of the Earth and ascend into
the atmosphere, it gets colder - until you pass a boundary
at which the temperature begins to increase again. There are
several such reversals in temperature gradient and the traditional
definition of atmospheric layers uses them to define the
layers of the atmosphere as `-spheres' with boundaries called `-pauses' \cite{Jursa1985}:

\begin{itemize}
\item The troposphere, between the ground and the tropopause
\item The stratosphere, between the tropopause and the stratopause (about 50 km)
\item The mesosphere, between the stratopause and the mesopause (about 85-90 km). Here 
CO2 cooling dominates solar heating.
\item The thermosphere, between the mesopause and the exobase (about 85 to 500 km, variable).
In the thermosphere the physical state is dominated by absorption of solar radiation;
the resulting ionized atoms have their own behaviour and the composition of the atmosphere
departs from the N2/O2 mix of the lower layers. The thermosphere region overlaps
(but has a definition which is not quite the same as)
the ionosphere, the region where ionized particles dominate the physics.  It includes
the LEO region where the ISS orbits.
\item The exosphere, beyond the exobase. Here the density is so low that the
atoms don't act like a gas.
\end{itemize}

Another relevant boundary is the turbopause, below which all the different molecules
have the same temperature, and above which they behave independently; below the
turbopause you are in the `homosphere' where everything is mixed; above it is the
'heterosphere' where everything acts independently. The turbopause is at about 100-120 km.
In 2009, press releases referencing a paper by Sangalli et al \cite{Sangalli2009} about
the Joule II rocket mission trumpeted a measurement of the `edge of space' at 118 km.
This was actually the height at which the motion of charged atoms (ions) becomes
dominated by the electromagnetic field rather than by winds in the neutral atmosphere;
it is likely a function of time and location and so their value for 2007 Alaska should
not be taken as a generic result.

The chemical composition of the atmosphere is largely constant up to the mesopause.
From a physical point of view, it is therefore reasonable to think of the
atmosphere proper as including the troposphere and stratosphere and (with some
qualification) the mesosphere, and
identifying the thermosphere and exosphere with the common idea of
'outer space'. Either the mesopause or the turbopause are reasonable choices
for a boundary, as the outermost physical atmospheric boundaries below the region
where most satellites orbit. It is true that each of these definitions varies in height by
10 km or more depending on solar activity, upper atmosphere dynamics and other factors.
The 1976 US Standard Atmosphere value for the mesopause is a constant 86 km;
Xu et al \cite{Xu2007} used observations with the SABER radiometer on the TIMED
satellite to study variations in the
mesosphere altitude. Their data suggest a mesosphere altitude of 97
$\pm$2 km for  equatorial and winter polar regions and 86$\pm$2 km for
summer polar regions; these  values are higher than I had assumed in
my 1994 discussion.

A reasonable alternative air/space boundary would be the base of the
mesosphere instead of its ceiling; or, one may consider the mesosphere
as neither air nor space. In 1976 Reijnen \cite{Reijnen1976} and Jager and Reijnen \cite{JagerReijnen1976}
introduced the idea of `mesospace' as an intermediate legal regime
between airspace and outer space; the mesosphere is a natural candidate
for mesospace. Oduntan \cite{Oduntan2003} suggested a buffer zone from 55 to 100
miles (88 to 160 km), apparently partly based on the existing incorrect
estimates of 150 km as the lowest orbital perigee.  In fact, as shown
below, 55 to 100 km would be a more suitable choice. Pelton \cite{Pelton2013} has
coined the term `protospace' or `the protozone' for the intermediate
region, which he defines as the 21 to 160 km range. In general, however,
the idea of mesospace has not yet gained general acceptance.

\subsection{Outer limits of the atmosphere and boundaries in deep space}\label{Sec302}

The true outer edge to the Earth's atmosphere, or a reasonable candidate
for it, is the magnetic shock front with the solar wind. The
magnetopause boundary forms a comet-shaped region, typically around the
height of geostationary orbit on the sunward side of Earth and extending
out to beyond the Earth-Sun L2 point. One can also consider the
gravitational boundary of the Earth-Moon system with respect to the Sun,
conventionally chosen to be the 1.5-million-km radius Hill sphere marked
by the Earth-Sun Lagrange points L1 and L2. While material within the
magnetosphere and/or the Hill sphere could be considered part of the
Earth's outer atmosphere, few would argue that this region is not
`space'. Rather, these boundaries may be used to distinguish space in
the Earth-Moon system from interplanetary space.

Indeed, one may usefully identify a number of different conventional regions 
in the region of space humans and their robots have explored, listed here
for convenience of reference. 

\begin{itemize}
\item The boundary between Low Earth Orbit (LEO) and Medium Earth Orbit (MEO),
is sometimes taken to be a 2 hour orbital period, which corresponds to an
altitude of 1682 km for equatorial orbits, but nowadays a round value of 2000 km is 
relatively standard, e.g. \cite{Johnson2010}.

\item The geosynchronous altitude, 35786 km above the equator \cite{Clarke1945}.

\item The Earth-Moon 1:4 resonance altitude EL1:4, 145688 km altitude. I introduce here
this boundary between `near-Earth space', where the effects of lunisolar perturbations
are minor and a simple Keplerian elliptical satellite orbit is a reasonable
approximation and `deep space', which I take to include both distant
Earth satellite orbits (such as that of the TESS satellite launched in 2018)
and lunar and planetary missions. For Earth satellite orbits in `deep space' the
lunar perturbations are large enough to make big changes in the orbital elements
on month-long timescales. As a practical matter, NORAD/JFSCC systematically
monitor orbits of near-Earth space spacecraft but do not attempt to monitor
deep-space Earth satellite orbits in a comprehensive way - this is left to
astronomers who accidentally pick up satellites in such orbits while searching
for asteroids. As will all these boundaries, one could reasonably make a different choice
here - a round altitude of 100,000 km, or a different resonance like EL1:3.
I propose EL1:4 by analogy with the Sun-Jupiter 1:4 resonance that is conventionally
taken to mark the inner edge of the asteroid belt (and thus the point inside which solar orbiting
objects can be considered as not strongly perturbed by Jupiter).

\item The Hill Sphere \cite{Hill1878}, bounded by the Earth-Sun Lagrange points, with a radius of 1.496 million km.
This is the conventional boundary between considering objects as orbiting Earth but perturbed by the Sun,
and considering objects as orbiting the Sun but (if close to the boundary) perturbed by the Earth.
Another choice here is the so-called `gravitational sphere of influence' or Laplace sphere \cite{Laplace1804},
which is at approximately
929,000 km radius; it is used in the method of patched conics.
In general the Hill sphere, which takes into account the
orbital angular momentum, better reflects the effective boundary
at which orbiting objects may be captured by or escape from the Earth-Moon system (e.g. \cite{VK2006},\cite{Araujo2008}).

\item The $\nu_6$ secular Sun-Jupiter-Saturn resonance which marks the
conventional inner edge of the asteroid belt at 2.06 astronomical units
(308 million km) from the Sun \cite{Tisserand1882}; it coincides with
the 1:4 Sun-Jupiter resonance \cite{Vakhidov1999} and asteroid orbits
near this resonance are unstable, soon perturbed to enter the inner
solar system. Although there is no generally agreed definition, this
location is a reasonable place to mark as the boundary between the inner and outer
solar system.

\item The outer edge of the Solar System itself is controversial. Plasma physicists associated with
studies by the Voyager probes  have made various estimates of the `heliopause' boundary between
the solar wind and the broader-scale flow of interstellar gas, for example
at 121.7 astronomical units ($1.8 \times 10^{10}$ km) \cite{WebberMcDonald2013}. However, dynamical
astronomers would point out that objects remain gravitationally bound to the Sun much
further out \cite{SMT84}, of order 200,000 astronomical units ($3\times 10^{13}$ km).

\end{itemize}

\section{Technological boundaries at the edge of space}\label{Sec4}

\subsection{The highest vehicles using aerodynamic lift}\label{Sec401}

On 1973 Jul 25 a modified Mig-25 designated the Ye-266 reached 36.2 km,
and on 1977 Aug 31 a Ye-266M reached 37.65 km. These altitude records
for non-rocket-powered airplanes were set
at the Soviet LII Gromov flight test center by test pilot Aleksandr Fedotov \cite{FAI2018};
they involved brief arcs to high altitude.
The highest steadily flying non-rocket plane was the remotely piloted Helios, which reached 29 km on
2001 Aug 14.  The highest crewed balloon reached 34.6 km in 1961.
However, uncrewed scientific research balloons reach over 40 km  on a
routine basis. A 1972 record of 51.8 km was exceeded in 2002 by the
54-meter-diameter BU60-1 balloon from the Japanese ISAS team, which
reached an altitude of 53 km \cite{Yamagami2003}. Despite the then-stated intent of the ISAS
team to reach 60 km, it appears that the technological limiting ceiling of vehicles which
require the atmosphere for lift is close to the stratopause at 50 km.
This sets a sensible lower limit for the boundary of space.

\subsection{The lowest quasi-circular orbits}\label{Sec402}

The much-cited FAI article \cite{FAI2004} about the Karman line continues:

\begin{quote}
{\it
Later in the same decade (or very early in the next; Soviet information
at the time was very scanty) the Soviet Union put in orbit an unmanned
satellite, in very low orbit, whose attitude was controlled by
aerodynamic forces. The real reason of such an experiment is not yet
known. It is known however that it successfully described a few orbits
just above the 100 km line (how much higher I do not know), but
collapsed rapidly shortly after he crossed, or got too much close to,
the 100 km. Karman line.
}
\end{quote}	

Soviet information was not that scanty even at the time; the author is
clearly referring to the well known Kosmos-149 satellite, which carried
an extendable structure used to stabilize it along the velocity vector.
This satellite, whose then-classified name was DS-MO No. 1, was launched
into a 245 x 285 km orbit, low enough for the drag stabilization to work
but much higher than needed to avoid catastrophic decay. It remained in
orbit from 1967 Mar 21 to Apr 7. The last US orbital data was on Apr 5,
at which time it was in a 201 km circular orbit.  But there are many
well documented cases of even lower altitude satellites.
Since the idea that 200 km is the low boundary for satellite orbits is so
widespread, I consider here a number of counterexamples.

In May 1976, the satellite GAMBIT Mission 4346 (1976-27A, US National
Reconnaissance Office) was tracked in an orbit with a perigee between
125 and 135 km for a full month; this is not unusual for this kind of
satellite, which performs frequent rocket burns to counteract decay. Its
apogee was  around 350 km. Empty rocket stages are frequently left in
low orbits of under 200 km and reenter after several days; the final
tracked orbit is often between 130 and 140 km.

From 2016 Aug 16-19, China's Lixing-1 satellite operated in a near-circular orbit
of 124 x 133 km for three days prior to reentry; this is the lowest
circular orbit ever sustained for multiple days.

In contrast, when the Space Shuttle lowered its perigee to 50 km as part
of the deorbit burn, it reenters within an orbit. Shuttle external
tanks, discarded at orbit insertion, often had perigees around 70-75 km
and in all cases did not complete their first orbit. 

Based on circular orbit data, 125 km is a conservative upper limit for 
the beginning of space.

\subsection{The lowest perigees for elliptical orbits}\label{Sec403}

A satellite in an elliptical orbit can survive a brief periapsis passage
at lower altitudes than the extended exposure of a circular orbit would
permit. Below I give examples of low perigee elliptical orbit satellites.
The air density increases rapidly, and so there is a limit below which even
a highly elliptical orbit satellite will be rapidly destroyed. This limit
turns out to be in the 80 to 90 km range except in very special cases.

Consider a satellite in an elliptical orbit whose perigee is around 80 km.
Are we to say that it is in space only for the higher parts of its orbit,
and that, for example, space law stops applying to it at each perigee passage?
The repetitive nature of an orbit makes this case different from the one-off transition
from the space to aviation environment during launch or reentry. I therefore
conclude that attempts to use `lowest circular orbit' to define the space boundary
are fundamentally misguided, and `lowest sustainable perigee' (for more than two
revolutions, say, of an elliptical orbit) is a more appropriate criterion.

Before considering specific examples, a detailed
discussion of the pitfalls in satellite perigee height calculations is warranted.

The Earth satellite catalog in widespread use is that currently
maintained by the US military, since the corresponding Russian catalog
is not publicly available and other sources (e.g. those from hobbyists)
are relatively incomplete. The catalog was begun in 1957 by the
Smithsonian Astrophysical Observatory \cite{SAOSR1}. The
North American Air Defense Command (NORAD) collaborated with SAO and
ultimately took over the catalog. Orbital data here and below are
obtained from the Two-Line Orbital Elements (TLE) issued by the US Joint
Space Force Component Command, \cite{JFSCC2018}, the current inheritor of
NORAD's space tracking responsibilities. For each satellite, there may
be several orbit determinations (`element sets', `TLE sets' or simply
`TLEs') per day. These data have been federated with spacecraft
historical information from the author's catalog of satellites \cite{McDowell2018}.

The TLEs provide mean motion and eccentricity of a fitted time-averaged
orbit, the `SGP4 mean elements' \cite{HootsRoehrich1980}, \cite{Vallado2006}.  It is common practice (notably in the official public satellite
catalog on space-track.org, or historically in the RAE Table Of Earth
Satellites and reports derived from it) to describe a satellite orbit by
quoting the perigee and apogee height of the SGP4 mean elements relative
to a fictitious 6378 km spherical Earth. To find the actual perigee
height of a satellite above the true surface of the Earth, one must
first apply the SGP4 theory to derive the osculating elements (or,
equivalently, state vector) at perigee. For an orbit with significant
eccentricity this perigee may be different from the SGP4 mean value by
of order ten kilometers. Next, the correction to the height above the
Earth ellipsoid rather than the spherical Earth model ranges up to 22 km
at the poles. I use the WGS-84 ellipsoid for calculations in this paper.

Low perigee TLEs are common for the final element set for a satellite,
but in some cases this may represent only the final orbit where perigee
is not survived. Single element sets are suspect; for example Kosmos-168
was tracked in a 52 x 386 km orbit on 1967 Jul 4, but data for
surrounding epochs make it clear that this was an erroneous solution
with the right period but wrong eccentricity. In particular, for
elliptical orbit satellites, as the orbital period decreases and perigee
drops below of order 100 km the sequence of mean element solutions show
increasing set-to-set noise and an increasing fraction of spurious fits.
In a review of an archive of 90 million TLE sets for 43000 
satellites I identified 50 satellites where the data are not severely
affected by these problems and where geodetic perigee heights
of less than 100 km were maintained over 2 or more complete revolutions of the Earth.

\begin{figure}[h]

\includegraphics[width=5.5in]{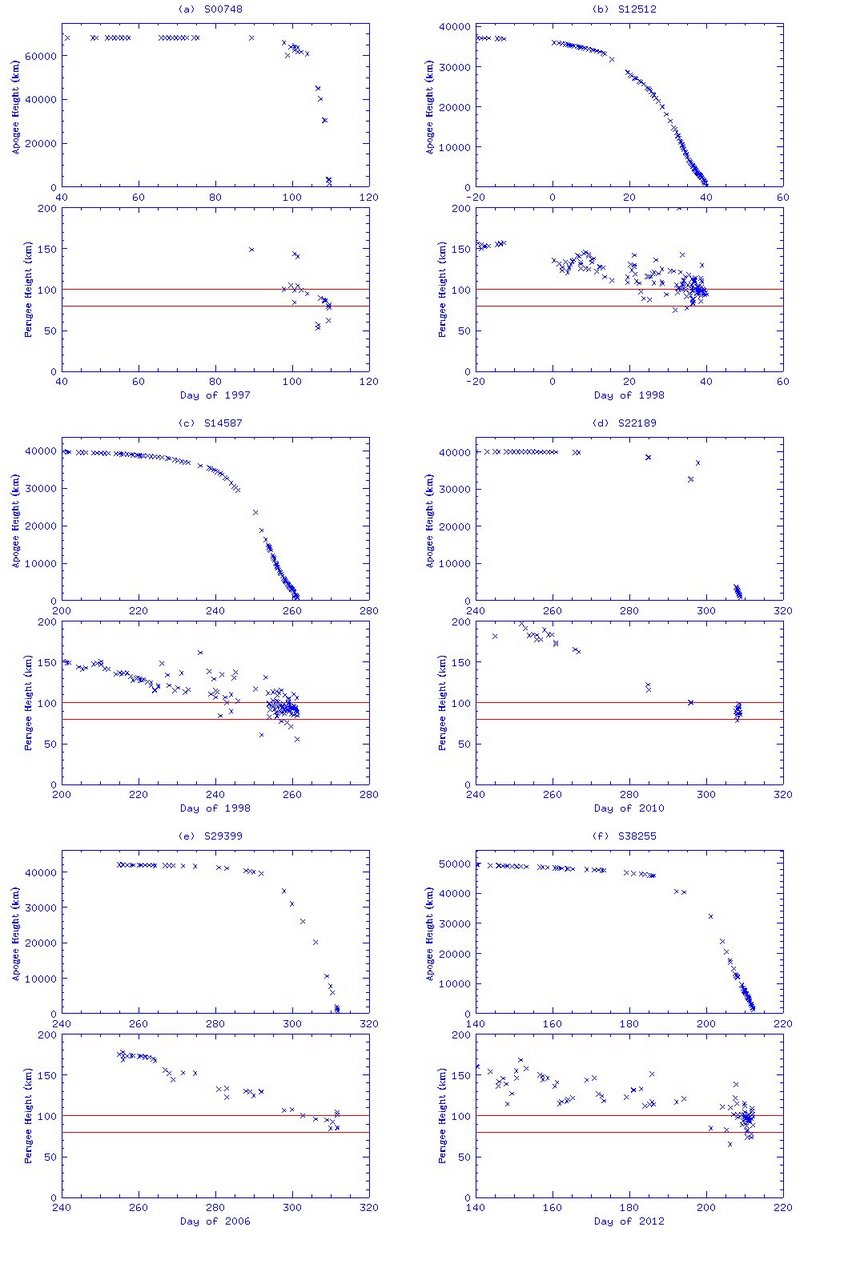}
\caption{
{\small Geodetic height of apogee and perigee versus time
for the decay of selected elliptical orbit satellites. Horizontal
lines at 80 and 100 km are superimposed on the perigee plots.
Despite noisy fits, these satellites appear to have survived 
multiple perigee passages below 100 km.
(a) Satellite 748 (1964-006B, Elektron 2, 2D No. 2); (b) Satellite 12512 (1981-30A, Molniya-3 No. 30);
(c) Satellite 14587 (1983-126A, Kosmos-1518, Oko 6022); (d) Satellite 22189 (1992-069A, Kosmos-2217, Oko 6059);
(e) Satellite 29399 (2006-038B, Chang Zheng 3A Y10 third stage rocket);
(f) Satellite 38255 (2012-019B, Centaur AV-031 rocket).}
}
\label{Fig1}
\end{figure}

A few illustrative examples are shown in Figure \ref{Fig1}. These include the Soviet
Elektron-4 satellite (SSN 748) which appears to have made 10 revolutions
with perigee at or below 85 km at the time of its reentry in 1997, and
the US Centaur AV-031 rocket (SSN 38255) which had perigee geodetic height
below 110 km for 4 days prior to reentry, and between 80 and 95 km for much of that time.

To summarize, the lowest possible sustained circular orbits are at of order
125 km altitude, but elliptical orbits with perigees at 100 km can survive
for long periods. In contrast, Earth satellites with perigees below 80 km are 
highly unlikely to complete their next orbit. It is noteworthy that meteors
(travelling much more quickly) usually disintegrate in the 70 to 100 km altitude range,
adding to the evidence that this is the region where the atmosphere becomes important.

\clearpage

\subsection{Air-space vehicles}\label{Sec404}

Goedhart  \cite{Goedhart1996}
correctly notes that during ascent a space vehicle
reaches 100 km altitudes in quite a short downrange distance, so that usually
crossing someone else's territory while in the atmosphere is not an issue, and normal
spacecraft landings are similar in this respect, but 
on reentry a winged spaceplane can be at 60 km altitudes or lower while traversing
long ground distances. The issue of airspace violations during spaceplane reentry
is therefore something to worry about. Goedhart correctly says that the limit 
of `aviation' and aerodynamic rather than ballistic phenomena is at 50 km (and infers that the Paris
and Chicago conventions apply up to this limit). He immediately 
contradicts himself by stating that `aeroplanes are already capable of flying
above the 50 km limit', which I suspect is a confusion between a true aeroplane and
rocketplanes like the X-15.  Several authors exhibit this confusion: like the later Shuttle,
the X-15 functions as an airplane (in fact, a glider) during descent through the lower atmosphere, and so
it looks like an airplane. But when it is above the lower atmosphere it uses rocket thrusters
to maneuver - it is then operating as a spacecraft despite its exterior appearance, and makes
no use of its aerodynamic surfaces. To use it as an example of a high altitude aircraft is
to miss the point.

\subsection{Other technological considerations}\label{Sec405}

Goedhart  \cite{Goedhart1996} also discusses what he calls the `biological theory', that at about 20 km
humans cannot survive unprotected due to low air pressure, but notes that clearly
air vehicles above this line are not considered to be in space.  

de Oliveira Bittencourt Neto \cite{Neto2015}
discusses the idea of `effective control' (you own the airspace you can control)
and points out that's a really bad idea as it leads to different boundaries of space above different
countries. Satellites would cross such boundaries every few minutes.

It is clear that for a definition of space to be useful and consistent with
the generally understood meaning of the term, it should be well above typical airplane
altitudes and should be globally uniform (this last constraint does not mean that one can't
adopt different globally-uniform definitions of space for different purposes and situations).

\section{The Effective Karman Line}\label{Sec6}

\subsection{Mathematical analysis of the Karman line}\label{Sec601}

The  `von Karman line' appears to be what mathematicians refer to as a `folk theorem',
arising out of a conference discussion but never formally published by him.
It was fleshed out in later publications, especially in the influential work of Haley (1963, \cite{Haley1963}) 
and there is some justification for calling it the `von Karman-Haley line'.  
\footnote{Note added in press: A recent paper by Gangale \cite{Gangale2018} untangles this complicated history in detail.}

von Karman's argument was that the space boundary should be defined where
forces due to orbital dynamics exceed aerodynamic forces. A rough order of magnitude
argument was used to show that this was at of order 100 km (as opposed to 10 km or 1000 km),
but in reality the von Karman criterion defines a line whose altitude varies with
position and time (because of variations in atmospheric density due to solar activity)
and with the lift coefficient of the spacecraft.

Haley (\cite{Haley1963}, p 78) extended the argument to satellite drag and
places the line at 84 km.  The strong association of the term `(von) Karman line' with
a definite 100 km value is a more recent development.

Satellite launch vehicles reach 100 km altitude in the first minutes of
flight, well before they have accelerated to orbital velocity; thus the
appropriate value of the parameter $f$ is less than one, and drag is
smaller and the gravity/drag force ratio correspondingly larger at a given altitude; hence the
effective Karman line is even lower in this phase or for suborbital missions.
I will consider only orbital flight in the following calculations.

I consider a spacecraft of mass m, cross sectional area A and lift and drag coefficients $C_L$ and $C_D$ travelling at velocity v, which
I'll later take to be the orbital Keplerian circular velocity $v_c$.
The spacecraft is travelling at geocentric radius r through atmosphere of density $\rho$
in the gravity field of the Earth whose mass is $M_E$.

The lift force is 
\begin{equation} 
 F = {1\over2}A C_L \rho v^2
\end{equation}
and the drag force is expressed my the same equation with a different coefficient $C_D$.

Following Haley I consider that drag forces are more relevant to Earth satellites, so instead consider
the ratio of drag force to weight (i.e. to gravitational pull).
Because atmospheric density changes by many orders of magnitude in a few tens of kilometers,
use of $C_D$ rather than $C_L$ does not change the final Karman Line location much, as we shall see below.

The ratio of gravitational force (weight W) to aerodynamic force (F) is 
\begin{equation}
 R = W/F = { mg \over {1\over 2} A C_D \rho v^2 } 
\end{equation}
where the local acceleration due to gravity is
\begin{equation} g = GM_E/r^2\end{equation}
and circular orbital velocity is
\begin{equation}
 v = \sqrt{ GM_E /r }
\end{equation}

We introduce the ballistic coefficient $B = C_DA/m$, which is essentially the specific drag, or drag
per unit mass (warning: some authors use the term ballistic coefficient for 1/B instead).
Then the above results can be simplified to
\begin{equation}
 R = { 2 \over B r \rho }
\end{equation}

When R is much greater than unity, orbital dynamics dominates aerodynamics
(and, per the original von Karman argument, lifting flight is not possible).

Because of the rapid change of density with height, R changes by
orders of magnitude in the range of interest. It is therefore convenient
to instead use the logarithm - I define 
\begin{equation}
   k(B,r,\rho) = \log_{10} R =  \log_{10} \left(  { 2  \over B r \rho  }\right)
\end{equation}
and call this logarithmic measure the {\bf Karman parameter}\footnote{Reijnen \cite{Reijnen1976} suggests a slightly different parameter,
the height at which drag reduces the height of a single 
circular orbit by 10 percent (and thus its radius from 
Earth center by about 0.2 percent). If $\delta$ is the fractional
change in the semi-major axis. Reijnen claims
\[
 \delta r = 2\pi B \rho r^2
\]
so $\delta = R$, hence Reijnen's criterion corresponds to a Karman
parameter k of -2.7 rather than 0.}.

\subsection{Reference ballistic coefficient and fiducial Karman parameter}\label{Sec602}

The Karman parameter is a function of position, atmospheric properties, and spacecraft ballistic coefficient B.
With a known atmospheric profile and value of B, one can derive $z_B(k)$, the
altitude at which k has a particular value. 
The value $k=0$ defines an effective Karman line height $z_B(0)$ at which aerodynamic
and gravitational forces balance.


Let us explore how much that line shifts around for a range of plausible values.
Typical values of $C_D$ are 2.0 to 2.4 for satellites, while
typical values of B are of order 0.006 to 0.05 $\mbox{m$^2$/kg}$
\cite{Bowman2002}, \cite{Saunders2012}. The International Space Station
has an average cross sectional area of 2040 sq m and a B of 0.010  $\mbox{m$^2$/kg}$, while the Planet Dove cubesats
have a cross-sectional area varying from 0.2 to 0.02 sq m and a B of 0.1 to 0.01 depending on flight attitude
\cite{Foster2015}.

However balloons and special high density satellites can have more
extreme values. The Echo balloon satellite had a high B of around 22
$\mbox{m$^2$/kg}$, while the LARES high density geodetic research
satellite had a record low B of around 0.001.


As an intermediate fiducial
value I adopt $B_0= 0.01 \mbox{m$^2$/kg}$ and define the {\bf fiducial Karman
parameter} $k_0$ as
\begin{equation}
 k_0(r,\rho) =  k(B_0,r,\rho)
\end{equation}
so that 
\begin{equation}
 k(B,r,\rho)  = k_0(r,\rho) -\log_{10}\left(B/B_0\right)
\end{equation}

We then define the function $z(x)$ as the geodetic altitude at which $k_0 = x$.
The usefulness of $k_0$ is that its values and the corresponding altitudes $z(k_0$) can be calculated for 
a given atmosphere independently of the satellite properties. One may then read off
the Karman line location for a particular satellite by determining which value of $k_0$ is
appropriate for its B.
For example if one considers a satellite
with a high B = 0.05, the Karman line $k=0$ corresponds to $k_0 = \log(5) = 0.70$.

Mathematically,
\begin{equation}
 z_B(x) = z(x+\log_{10}\left(B/B_0\right))
\end{equation}
with the Karman line at x=0.

\begin{figure}[h]
\includegraphics[width=5.5in]{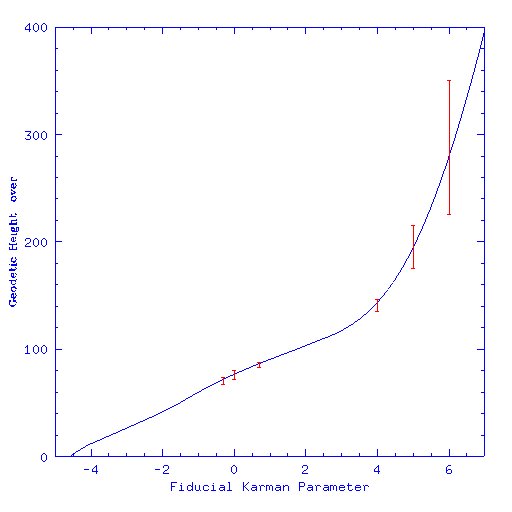}
\caption{Fiducial Karman parameter $k_0$ versus geodetic altitude for US Standard Atmosphere 1976.
The red error bars summarize the results of figures \ref{Fig3} and \ref{Fig4}, 
indicating the range of variation in the altitude for a given value of $k_0$
found in runs of the NRL atmosphere model for different dates, latitudes and longitudes.}
\label{Fig2}
\end{figure}

\subsection{Numerical evaluation of the effective Karman line}\label{Sec603}

I now derive the geodetic altitude z for various values of the fiducial Karman parameter
and consider how the Karman parameter changes with location and time.

A useful reference atmosphere is the 
US Standard Atmosphere 1976 (hereafter USSA76) \cite{USSA1976}, which is a single, fixed,
atmosphere model. Figure \ref{Fig2} shows the altitude as a function
of fiducial Karman parameter for this atmosphere.
However, in practice the atmospheric density
at a given altitude varies with longitude and latitude, and also with solar activity
as the atmosphere is heated by the solar flux. 

To understand the effect of these variations on the $k$ I
ran models for assorted times and geographical locations using a code
which implements the NRL MSISE-00 atmospheric model \cite{Picone+2002}.
Atmospheres were calculated at 10 day intervals from Jan 1960 to Jan
2020 to fully sample several solar cycles. At each  selected day 
atmospheres were calcualted for 0, 6, 12 and 18h GMT at  four latitudes
(80S, 0N, 45N and 80N) and four longitudes (0, 90, 180, 270E). For each
epoch, actual or (for future dates) predicted solar activity levels from
the Celestrak space weather archive \cite{Kelso2017}, \cite{ValladoKelso2013} were used.

\begin{figure}[h]
\includegraphics[width=5.5in]{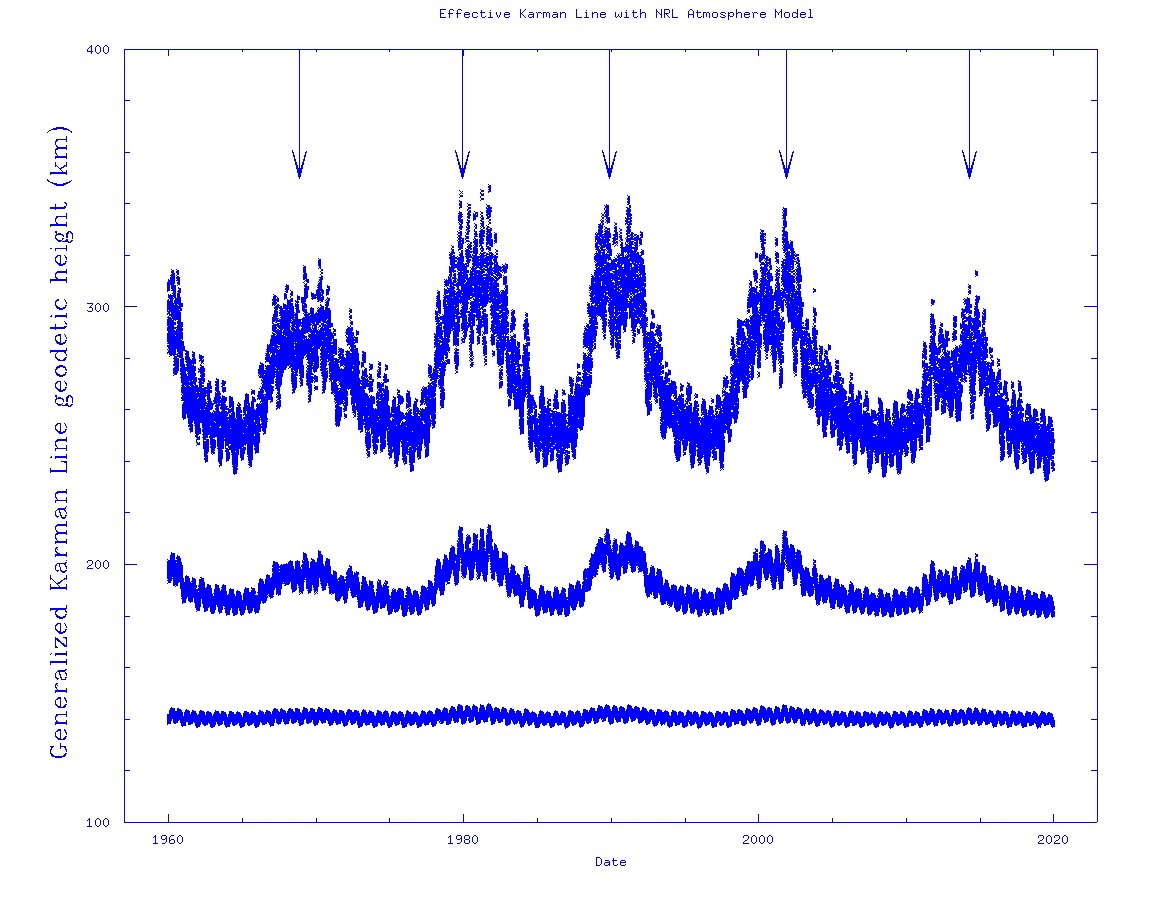}
\caption{ Curves showing z(4) (lowest), z(5) and z(6) (highest)
as a function of time, showing that
the effects of the solar cycle are more important at high Karman
parameter. These integrations are for NRL atmosphere models evaluated
at 45 deg N, but curves for
other latitudes are similar. Arrows indicate dates of solar maxima.}
\label{Fig3}
\end{figure}

In Figure \ref{Fig3} I show the $z(k_0)$ lines for high values of $k_0$, namely
$k_0$ = 4,5,6. There are high amplitude variations in the $k_0=6$ Karman
line correlated with the solar cycle, reflecting the well-known
sensitivity of atmospheric density to solar flux at the corresponding
altitudes, above the mesopause.  Note, however, the much reduced
amplitude of the variations at lower altitudes and Karman parameter
values. The USSA76 model values are in all cases within the range of the NRL model
variations; the range of the variations at each modelled parameter value is summarized 
as an error bar on the USSA76 curve in Figure \ref{Fig2}.

\begin{figure}[h]
\includegraphics[width=3.2in]{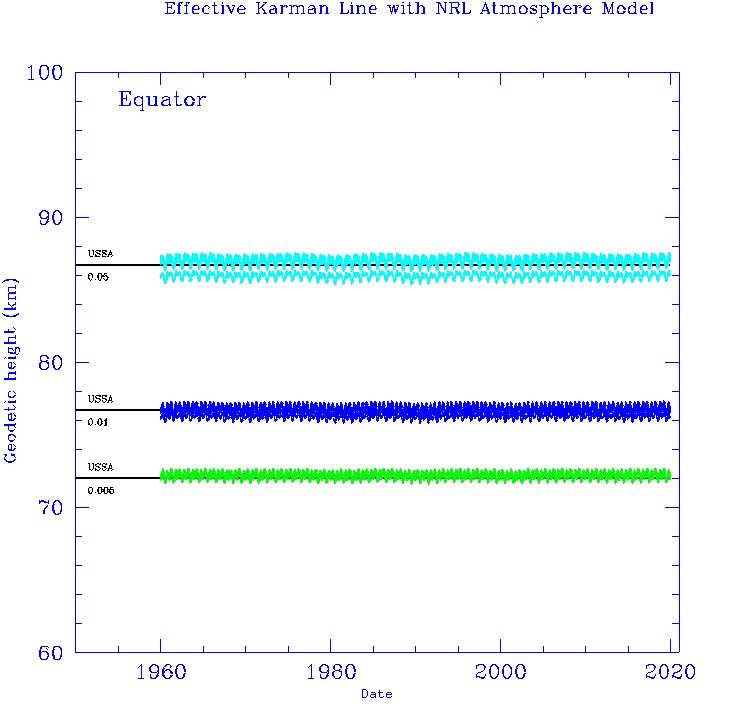}
\includegraphics[width=3.2in]{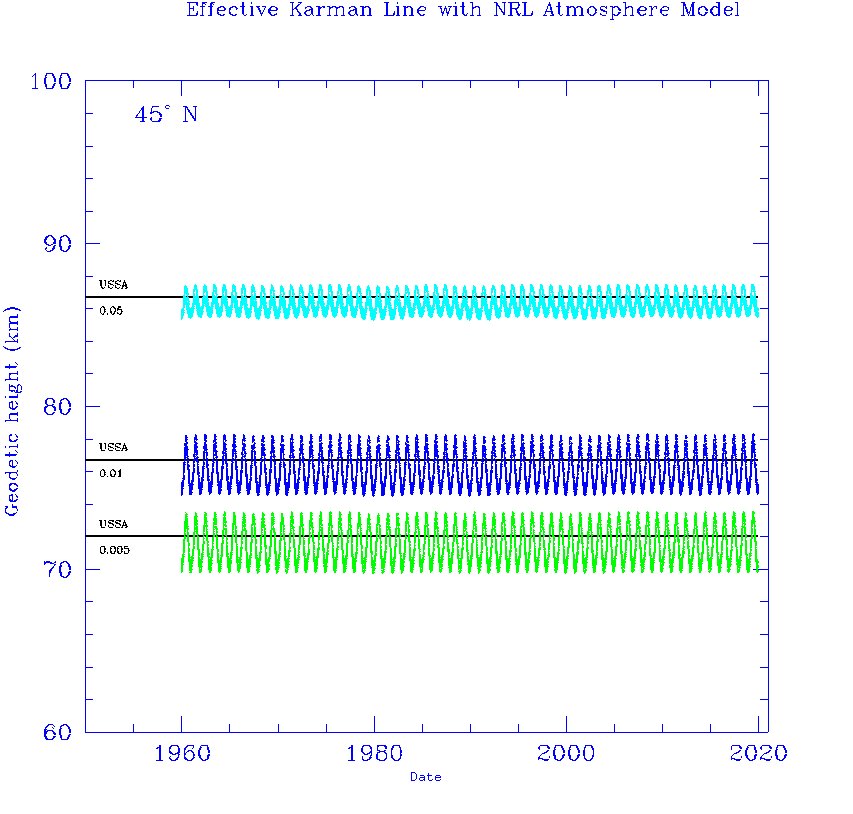}
\includegraphics[width=3.2in]{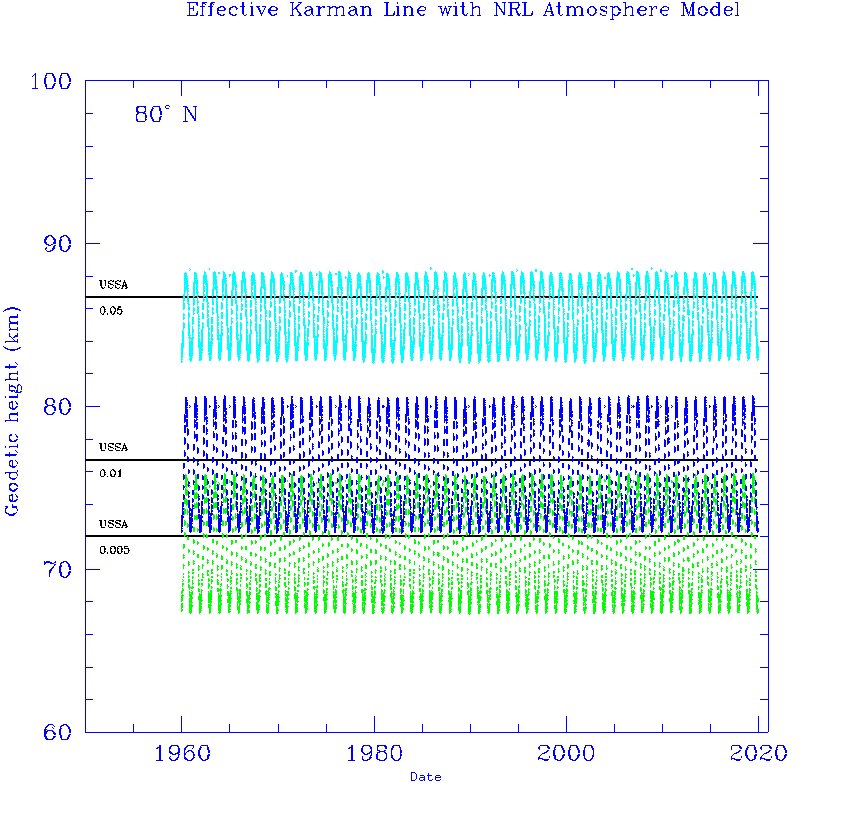}
\includegraphics[width=3.2in]{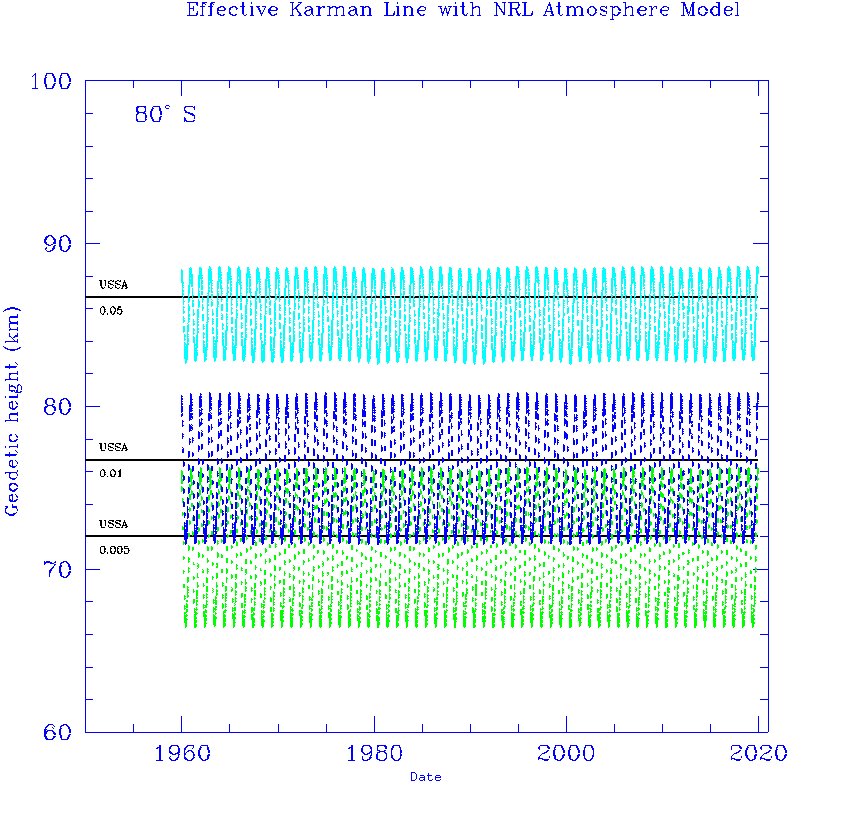}

\caption{Curves for z(-0.3), z(0.0), z(0.7), corresponding
to the effective Karman line for $B = 0.005, 0.01, 0.05 \mbox{m$^{2}$ kg$^{-1}$}$
respectively. Each plot gives calculations for a different latitude; there is
less atmospheric variation at intermediate latitudes. At these low altitudes
the effects of the solar cycle are minimal.}
\label{Fig4}
\end{figure}

In particular we are interested in fiducial Karman parameters $k_0$
equal to -0.3,0.0,+0.7, corresponding to effective Karman lines $k=0$
covering the typical range of B for satellites discussed above. 
The USSA1976 reference model gives
z(-0.3)=72.0 km, z(0.0)= 76.7 km, and z(0.7)=86.7 km.
In Figure \ref{Fig4} I show corresponding NRL model atmosphere calculations as a function of time.
Seasonal variations are most prominent at polar latitudes but their
amplitude is only a few kilometres. The fact that actual historical solar flux values
for each date were used confirm that irregular solar flares do not affect the result.

In all cases the effective Karman lines calculated from the NRL atmospheres remain within 
5 km or so of their USSA values, and the overall range of the data (for a factor of 10 in ballistic
coefficient) is from 66 to 88 km. In other words, the region where 
aerodynamic forces transition from dominant to negligible is relatively
well defined despite typical variations in satellite and atmosphere properties (at least
to the extent that this atmosphere model reflects reality).
The range of ballistic coefficients considered here is comparable to the difference
between typical lift and drag coefficients, and so this conclusion still holds if the original lift-based
Karman criterion is preferred.

It is true that for satellites with extreme properties the results do change - for a balloon satellite
like Echo, the effective Karman line z(1) is around 140 km. For a very dense satellite such as LARES,
the effective Karman line z(-1) is around 60 km. 

\clearpage

It is undesirable to have a definition that will change with improving
technology, so one might argue that the correct way to define space is
to pick the lowest altitude at which {\bf any} satellite can remain in
orbit, and thus the lowest ballistic coefficent possible should be
adopted - a ten-meter-diameter solid sphere of pure osmium, perhaps,
which would have B of $8\times 10^{-6}\mbox{m$^2$/kg}$ and an effective
Karman line of z(-4) at the tropopause. In practice z(0) seems a more realistic limit
for finite orbital lifetime (see previous section). The few high density
satellites at low orbital altitudes (e.g. LOADS 2, B of around 0.002 \mbox{m$^2$/kg})
have reentered when their perigees were around 120 km.

We can summarize the results of this section by saying that for a vehicle of typical
ballistic coefficient, z(0) represents the altitude at which gravity will exceed
aerodynamic forces for any object in steady flight at that altitude
(since such flight must always be at or less than the Keplerian circular velocity
for that altitude). That altitude lies in the 70 to 90 km range, and 100 km is
always too high.



\subsection{Extension of the Karman argument to elliptical orbits}\label{Sec604}

In section \ref{Sec403} I showed that elliptical orbit satellites can survive lower altitudes
than circular orbit ones. I now derive the ratio of gravitational to aerodynamic
force at the perigee of an elliptical orbit.

For an orbit of eccentricity e, perigee velocity is related to the circular velocity for
that altitude by
\begin{equation}
 v = v_c \sqrt{1+e}
\end{equation}
and so
\begin{equation}
 R = { 2 \over B r \rho (1+e) },
\end{equation}
i.e. the ratio is lowered by a factor of one plus the orbit eccentricity, so the effective
Karman line is actually somewhat higher. The low-perigee satellite is indeed drag-dominated
near perigee, causing rapid reduction of apogee and consquently reduction of the eccentricity,
so that close to reentry the Karman ratio tends to the circular value.

\section{Conclusion}

I have shown that for a typical satellite ballistic coefficient the effective Karman line
is close to (within 10 km of) 80 km independent of solar and atmospheric conditions,
rather than the currently popular 100 km value; and
that historical orbital data for actual artificial satellites confirms that orbiting
objects can survive multiple perigees at altitudes around 80 to 90 km. 
This altitude range is consistent with the highest physical boundary in the atmosphere,
i.e. the mesopause, and with the 50-mile `astronaut wings' boundary suggested by the United
States during the first years of the Space Age. 

On the basis of these physical, technological and historical arguments, I therefore suggest
that a value of 80 km is a more suitable choice to use as the canonical lower `edge of space'
in circumstances where such a dividing line between atmosphere and space is desired.



\vskip 0.1in
\subsection*{Acknowledgements}

I thank Brian Weeden, Joe Pelton and Isabelle Rongier for critical comments on
an earlier version of this paper. They will still disagree with its conclusions.
I also thank Martin Elvis for his careful reading of the manuscript.

This research did not receive any specific grant from funding agencies in the
public, commercial or not-for-profit sectors.
  
\subsection*{References}

\bibliography{bib1}

\begin{thebibliography}{44}
\providecommand{\natexlab}[1]{#1}
\providecommand{\url}[1]{\texttt{#1}}
\expandafter\ifx\csname urlstyle\endcsname\relax
  \providecommand{\doi}[1]{doi: #1}\else
  \providecommand{\doi}{doi: \begingroup \urlstyle{rm}\Url}\fi

\bibitem[{Jastrow}(1957)]{Jastrow1957}
R.~{Jastrow}.
\newblock {Definition of Air Space}.
\newblock In \emph{Proc. 1st Coll on Law of Outer Space (IISL, the Hague).},
  1957.

\bibitem[{Goedhart}(1996)]{Goedhart1996}
R.~{Goedhart}.
\newblock \emph{{The Never Ending Dispute: Delimitation of Air Space and Outer
  Space}}.
\newblock Editions Frontieres, 1996.

\bibitem[{Haley}(1963)]{Haley1963}
A.G. {Haley}.
\newblock \emph{{Space Law and Government}}.
\newblock New York: Appleton-Century-Crofts, 1963.

\bibitem[{Harris} and {Harris}(2006)]{HarrisHarris2006}
A.~{Harris} and R.~{Harris}.
\newblock {The need for air space and outer space demarcation}.
\newblock \emph{Space Policy}, 22:\penalty0 3, 2006.

\bibitem[{Monahan}(2008)]{Monahan2008}
R.~{Monahan}.
\newblock \emph{{The Sky's The Limit?}}
\newblock PhD thesis, Durham University, 2008.

\bibitem[{de Oliveira Bittencourt Neto}(2015)]{Neto2015}
O.~{de Oliveira Bittencourt Neto}.
\newblock \emph{{Defining the Limits of Outer Space for Regulatory Purposes}}.
\newblock Springer/ISU, 2015.

\bibitem[{Hansen}(2015)]{Hansen2015}
R.~{Hansen}.
\newblock {An Inductive Approach to the Air-Space Boundary Question}.
\newblock Working Paper No. 148, KU Leuven Center for Global Governances
  Studies, 2015.

\bibitem[on~the Peaceful Uses~of Outer~Space)(2002)]{COPUOS2002}
COPUOS (UN GA~Committee on~the Peaceful Uses~of Outer~Space).
\newblock {Historical summary on the consideration of the question of the
  definition and delimitation of outer space}.
\newblock Technical Report A/AC.105/769, United Nations, 2002.

\bibitem[{McDougal} and {Lipson}(1958)]{McDougalLipson1958}
M.S. {McDougal} and L.~{Lipson}.
\newblock {Perspectives for a Law of Outer Space}.
\newblock Faculty Scholarship Series Paper 2618, {Yale Law School}, 1958.

\bibitem[{Thompson}(1992)]{Thompson1992}
M.O. {Thompson}.
\newblock \emph{{At The Edge of Space: The X-15 Flight Program}}.
\newblock Washington, D.C.: Smithsonian Institution Press, 1992.

\bibitem[{McDowell}({1994})]{McDowell1994}
J.C. {McDowell}.
\newblock {The X-15 Spaceplane}.
\newblock \emph{Quest}, Spring {1994}.

\bibitem[{Sanz Fernadez de Cordoba}(1994)]{FAI2004}
S.~{Sanz Fernadez de Cordoba}.
\newblock {100 km Altitude Boundary for Astronautics}, 1994.
\newblock URL
  \url{http://www.fai.org/icare-records/100km-altitude-boundary-for-astronautics}.
\newblock [Online; accessed 9-May-2018; accessed 1-Jul2004 at
  http://www.fai.org/astronautics/100km.asp].

\bibitem[{Jursa}(1985)]{Jursa1985}
A.S. {Jursa}.
\newblock \emph{{Handbook of Geophysics and the Space Environment}}.
\newblock {Air Force Geophysics Laboratory, United States Air Force}, 1985.

\bibitem[{Sangalli} et~al.(2009){Sangalli}, {Knudsen}, {Larsen}, {Zhan},
  {Pfaff}, and {Rowland}]{Sangalli2009}
L.~{Sangalli}, D.~J. {Knudsen}, M.~F. {Larsen}, T.~{Zhan}, R.~F. {Pfaff}, and
  D.~{Rowland}.
\newblock {Rocket-based measurements of ion velocity, neutral wind, and
  electric field in the collisional transition region of the auroral
  ionosphere}.
\newblock \emph{Journal of Geophysical Research (Space Physics)}, 114:\penalty0
  A04306, April 2009.
\newblock \doi{10.1029/2008JA013757}.

\bibitem[{Xu} et~al.(2007){Xu}, {Smith}, {Yuan}, {Liu}, {Wu}, {Mlynczak}, and
  {Russell}]{Xu2007}
J.~{Xu}, A.~K. {Smith}, W.~{Yuan}, H.-L. {Liu}, Q.~{Wu}, M.~G. {Mlynczak}, and
  J.~M. {Russell}.
\newblock {Global structure and long-term variations of zonal mean temperature
  observed by TIMED/SABER}.
\newblock \emph{Journal of Geophysical Research (Atmospheres)}, 112\penalty0
  (D11):\penalty0 D24106, December 2007.
\newblock \doi{10.1029/2007JD008546}.

\bibitem[{Reijnen}(1976)]{Reijnen1976}
G.C.M. {Reijnen}.
\newblock \emph{{Legal Aspects of Outer Space}}.
\newblock PhD thesis, Univ. of Utrecht, 1976.

\bibitem[{{Jager}, C.} and {{Reijnen}, G.C.M.}(1976)]{JagerReijnen1976}
{{Jager}, C.} and {{Reijnen}, G.C.M.}
\newblock {Mesospace}.
\newblock In \emph{Proc. 18th Coll. IISL/IAF}, 1976.

\bibitem[{Oduntan}(2003)]{Oduntan2003}
G.~{Oduntan}.
\newblock {The Never Ending Dispute: legal theories on the spatial demarcation
  boundary plane between airspace and outer space}.
\newblock \emph{Hertfordshire Law Journal}, 1:\penalty0 64--84, 2003.

\bibitem[{Pelton}(2013)]{Pelton2013}
J.~{Pelton}.
\newblock {Beyond the Protozone}.
\newblock In \emph{ABA Forum on Air and Space Law, Wash. DC, Jun 6, 2013},
  2013.

\bibitem[{Johnson}(2010)]{Johnson2010}
N.L. {Johnson}.
\newblock {Medium Earth Orbits: Is There A Need For a Third Protected Region?}
\newblock In \emph{{61st International Astronautical Congress, 27 Sep - 1 Oct
  2010}}, 2010.
\newblock URL
  \url{https://ntrs.nasa.gov/archive/nasa/casi.ntrs.nasa.gov/20100007939.pdf}.

\bibitem[{Clarke}(1945)]{Clarke1945}
A.C. {Clarke}.
\newblock {Extra-Terrestrial Relays}.
\newblock \emph{{Wireless World}}, pages 305--308, October 1945.

\bibitem[{Hill}(1878)]{Hill1878}
G.W. {Hill}.
\newblock {Researches in the Lunar Theory}.
\newblock \emph{{American Journal of Mathematics}}, 1\penalty0 (1):\penalty0
  5--26, 1878.
\newblock \doi{10.2307/2369430}.

\bibitem[{Laplace}(1804)]{Laplace1804}
P-S. {Laplace}.
\newblock {Des perturbations que les com\`{e}tes \'{e}prouvent lorsqu'elles
  approchent tr\`{e}s-pr\`{e}s des plan\`{e}tes}.
\newblock In \emph{{Trait\'{e} de m\'{e}canique c\'{e}leste, Tome IV, Livre IX,
  Ch. II}}. {A Paris : Chez Courcier}, 1804.

\bibitem[{Valtonen} and {Karttunen}(2006)]{VK2006}
M.~{Valtonen} and H.~{Karttunen}.
\newblock \emph{{The Three-Body Problem}}.
\newblock {Cambridge, UK: Cambridge University Press}, 2006.

\bibitem[{Araujo} et~al.(2008){Araujo}, {Winter}, {Prado}, and {Vieira
  Martins}]{Araujo2008}
R.~A.~N. {Araujo}, O.~C. {Winter}, A.~F.~B.~A. {Prado}, and R.~{Vieira
  Martins}.
\newblock {Sphere of influence and gravitational capture radius: a dynamical
  approach}.
\newblock \emph{Mon. Not. R. Astron. Soc.}, 391:\penalty0 675--684, December
  2008.
\newblock \doi{10.1111/j.1365-2966.2008.13833.x}.

\bibitem[{Tisserand}(1882)]{Tisserand1882}
F.~{Tisserand}.
\newblock {Memoire sur les mouvements seculaires des plans des orbites de trois
  plan\`{e}tes}.
\newblock \emph{Annales de l'Observatoire de Paris}, 16:\penalty0 E.1--E.57,
  1882.

\bibitem[{Vakhidov}(1999)]{Vakhidov1999}
A.~A. {Vakhidov}.
\newblock {Asteroid orbits near the 4:1 resonance with Jupiter.}
\newblock \emph{Baltic Astronomy}, 8:\penalty0 425--441, 1999.

\bibitem[{Webber} and {McDonald}(2013)]{WebberMcDonald2013}
W.~R. {Webber} and F.~B. {McDonald}.
\newblock {Recent Voyager 1 data indicate that on 25 August 2012 at a distance
  of 121.7 AU from the Sun, sudden and unprecedented intensity changes were
  observed in anomalous and galactic cosmic rays}.
\newblock \emph{Geophysics Research Letters}, 40:\penalty0 1665--1668, May
  2013.
\newblock \doi{10.1002/grl.50383}.

\bibitem[{Smoluchowski} and {Torbett}(1984)]{SMT84}
R.~{Smoluchowski} and M.~{Torbett}.
\newblock {The boundary of the solar system}.
\newblock \emph{Nature}, 311:\penalty0 38, September 1984.
\newblock \doi{10.1038/311038a0}.

\bibitem[{Federation Aeronautique Internationale}(2018 \hfill)]{FAI2018}
{Federation Aeronautique Internationale}.
\newblock {Records}, 2018 \hfill.
\newblock URL \url{https://fai.org/records?record=fedotov}.
\newblock [Online; accessed 9-May-2018].

\bibitem[{Yamagami}(2003)]{Yamagami2003}
T.~{Yamagami}.
\newblock {Research on Balloons to Float Over 50 km Altitude}, 2003.
\newblock URL \url{http://www.isas.jaxa.jp/e/special/2003/yamagami/03.shtml}.
\newblock [Online; accessed 9-May-2018].

\bibitem[{Schilling} and {Sterne}(1957)]{SAOSR1}
G.~F. {Schilling} and T.~E. {Sterne}.
\newblock {Preliminary Orbit Information for U.S.S.R. Satellites 1957
  {$\alpha$}1 and {$\alpha$}2}.
\newblock \emph{SAO Special Report}, 1, October 1957.

\bibitem[(JFSCC)(2018)]{JFSCC2018}
Joint Force Space Component~Command (JFSCC).
\newblock {Space-Track.Org}, 2018.
\newblock URL \url{http://www.space-track.org}.
\newblock [Online; accessed 1-May-2018. Web site developed under contract by
  SAIC].

\bibitem[{McDowell}({2018, in prep.})]{McDowell2018}
J.C. {McDowell}.
\newblock {General Catalog of Space Objects}.
\newblock {2018, in prep.}
\newblock Preliminary version available at
  https://planet4589.org/space/log/satcat.txt.

\bibitem[{Hoots} and {Roehrich}(1980)]{HootsRoehrich1980}
F.R. {Hoots} and R.L. {Roehrich}.
\newblock {SPACETRACK REPORT No. 3, Models for Propagation of NORAD Element
  Sets}.
\newblock Technical report, US Air Force Aerospace Defense Command, Colorado
  Springs, Colorado., 1980.

\bibitem[{Vallado} et~al.(2006){Vallado}, {Crawford}, {Hujsak}, and
  {Kelso}]{Vallado2006}
D.~{Vallado}, P.~{Crawford}, R.~{Hujsak}, and T.S. {Kelso}.
\newblock {Revisiting Spacetrack Report \#3, Paper AIAA 2006-6753}.
\newblock In \emph{AIAA Astrodynamics Specialist Conference}, 2006.

\bibitem[{Gangale}(2017)]{Gangale2018}
T.~{Gangale}.
\newblock {The Non Karman Line: An Urban Legend of the Space Age}.
\newblock \emph{{J. Space Law}}, 41\penalty0 (2), 2017.

\bibitem[{Bowman}(2002)]{Bowman2002}
B.~{Bowman}.
\newblock {True Satellite Ballistic Coefficient Determination for HASDM}.
\newblock In \emph{AIAA/AAS Astrodynamics Specialist Conference and Exhibit},
  2002.

\bibitem[{Saunders} et~al.(2012){Saunders}, {Swinerd}, and
  {Lewis}]{Saunders2012}
A.~{Saunders}, G.~G. {Swinerd}, and H.~G. {Lewis}.
\newblock {Deriving Accurate Satellite Ballistic Coefficients from Two-Line
  Element Data}.
\newblock \emph{Journal of Spacecraft and Rockets}, 49:\penalty0 175--184,
  January 2012.
\newblock \doi{10.2514/1.A32023}.

\bibitem[{Foster} et~al.(2015){Foster}, {Hallam}, and {Mason}]{Foster2015}
C.~{Foster}, H.~{Hallam}, and J.~{Mason}.
\newblock {Orbit Determination and Differential-drag Control of Planet Labs
  Cubesat Constellations; AAS Paper 15-524}.
\newblock \emph{ArXiv e-prints}, September 2015.

\bibitem[{National Oceanic and Atmospheric Administration}(1976)]{USSA1976}
{National Oceanic and Atmospheric Administration}.
\newblock \emph{{U.S. Standard Atmosphere 1976}}.
\newblock Washington, D.C.: U.S. Govt. Print. Off., 1976.

\bibitem[{Picone} et~al.(2002){Picone}, {Hedin}, {Drob}, and
  {Aikin}]{Picone+2002}
J.~M. {Picone}, A.~E. {Hedin}, D.~P. {Drob}, and A.~C. {Aikin}.
\newblock {NRLMSISE-00 empirical model of the atmosphere: Statistical
  comparisons and scientific issues}.
\newblock \emph{Journal of Geophysical Research (Space Physics)}, 107:\penalty0
  1468, December 2002.
\newblock \doi{10.1029/2002JA009430}.

\bibitem[{Kelso}(2018 \hfill)]{Kelso2017}
T.S. {Kelso}.
\newblock {Celestrak, Space Weather Data}, 2018 \hfill.
\newblock URL \url{https://celestrak.com/SpaceData/sw19571001.txt}.
\newblock [Online; accessed 9-May-2018].

\bibitem[{Vallado} and {Kelso}(2013)]{ValladoKelso2013}
D.A. {Vallado} and T.S. {Kelso}.
\newblock {Earth Orientation Parameter and Space Weather Data for Flight
  Operations}.
\newblock In \emph{{23rd AAS/AIAA Space Flight Mechanics Meeting, Kauai, HI}},
  2013.

\end{thebibliography}

\end{document}